\documentclass[12pt]{article}

\usepackage{}
\usepackage{natbib}

\newcommand{\ex}[1]{\langle #1 \rangle}

\newcommand{\mc}[1]{\mathcal{#1}}

\usepackage{enumitem}

\newcommand{\ket}[1]{|#1\rangle}

\newcommand{\bkt}[2]{\langle#1|#2\rangle}

\newtheorem{theorem}{Theorem}

\title{Subjectivists about Quantum Probabilities Should be Realists about Quantum States}
\author{Wayne C. Myrvold \\
 Department of Philosophy \\ The University of Western Ontario \\ {wmyrvold@uwo.ca}}
 \date{Forthcoming in Meir Hemmo and Orly Shenker, eds., \emph{Quantum, Probability, Logic: The Work and Influence of Itamar Pitowsky} (Springer Nature, 2020).}

\begin{document}
\maketitle
\begin{abstract}
{There is a significant body of literature, which includes Itamar Pitowksy's ``Betting on Outcomes of Measurements,''  that  sheds light on the structure of quantum mechanics, and the ways in which it differs from classical mechanics, by casting the theory in terms of agents' bets on the outcomes of experiments. Though this approach, by itself, is neutral as to the ontological status of quantum observables and quantum states, some, notably those who adopt the label ``QBism'' for their views, take this approach as providing incentive to conclude that quantum states represent nothing in physical reality, but, rather, merely encode an agent's beliefs. In this chapter, I will argue that the arguments for realism about quantum states go through when the probabilities involved are taken to be subjective, if the conclusion is about the agent's beliefs: an agent whose credences conform to quantum probabilities should believe that preparation procedures with which she associates distinct pure quantum states produce  distinct states of reality.  The conclusion can be avoided only by stipulation of limitations on the agent's theorizing about the world, limitations that are not warranted by the empirical success of quantum mechanics or any other empirical considerations.  Subjectivists about quantum probabilities should be realists about quantum states.}
\end{abstract}
\section{Introduction}
There is a conception of probability, which often goes by the name of \emph{subjective probability}, that  takes probability to have to do with an agent's degrees of belief.  An important step in the development of this conception was its integration with decision theory, pioneered by Ramsey and de Finetti, and developed more fully by Savage.  On this view, an agent's preferences between acts (often illustrated by choices of bets to make or accept) are taken to be indications of the agent's degrees of belief in various propositions.

It is an attractive idea to extend this conception to quantum probabilities. One systematic development of this idea can be found in Pitowsky's ``Betting on the outcomes of measurements:
a Bayesian theory of quantum probability'' \citep{PitowskyQBet}.  The idea also forms the basis of the position (or family of positions) known as QBism \citep{CFSQB,CFS2007,FSQBC,FMSQB,FuchsNotwithstanding,QBHero}.  Proponents of such views tend to also take quantum states to be \emph{nothing more} than codifications of an agent's beliefs.  But the argument for this conclusion is never spelled out with sufficient clarity.

On the other hand, we find arguments---of which the theorem of Pusey, Barrett, and Rudolph (\citeyear{PBR}) is the best known---for the opposite conclusion, namely, that quantum states should be taken to represent elements of physical reality.  If interpreting quantum probabilities as subjective leads one inexorably to rejecting an ontic view of quantum states, then there must be something about the reasoning behind these arguments that doesn't go through on a subjective reading of probability.  In this chapter, I will argue that this is not the case.  Suitably interpreted, a version of the PBR argument leads to the conclusion that  an agent who takes quantum mechanics as a guide for setting her subjective degrees of belief about the outcomes of future experiments ought to believe that preparations with which she associates distinct pure states result in ontically distinct physical states of affairs (this terminology will be explained in section \ref{Fram}, below).  Note that this conclusion is \emph{entirely about the agent's beliefs}.  We are not presuming, as part of the argument, that the agent's beliefs in any way reflect physical reality.  That is a matter for the agent's own judgment (though, of course, there is an implication that if \emph{you} take quantum  mechanics as a guide for setting your degrees of belief about the outcomes of future experiments, you should take distinct pure states to be ontically distinct).

The argument, of course, involves assumptions about the agent's credences. It is also assumed that the agent may entertain theories about the way the world is, and have degrees of belief in such theories.  It is, of course, \emph{logically possible} to reject all such theorizing.  If such a move is proposed in connection with quantum theory, then the question arises whether the empirical evidence that led the community of physicists to accept quantum theory over classical theory provides any incentive for the move.  I will argue that we have no reason whatsoever to make such a move.  Anyone is, of course, free to make such a move without reason, as a free choice.  Dissuasion of someone so inclined will not be the topic of this chapter; what concerns us here is the evidential situation.

\section{Credence and action: some preliminary considerations}  In this section I would like to bring to the reader's mind some considerations of the sort that may be called ``pre-theoretical''---the sort of considerations that tend to be taken for granted, prior to any physical theorizing. They are, for that reason, to be thought of as neither particularly classical nor quantum.  I do not take myself to be saying anything new, radical, or contentious.  Of course, one may start a scientific investigation with certain presuppositions about the way the world is, and end up concluding, as a result of one's investigations, that some of those presuppositions are false. If any readers are inclined to dispute some assertion made in this section, then I will ask that reader to provide evidence for the falsity of the assertion.  I hope that it goes without saying that the mere  fact that some proposition leads to an unwelcome conclusion does not suffice as evidence that the proposition is false.

Consider the following situation.  You are attending a lecture, and you are taking notes on a laptop.  This process involves a sequence of choices: you are choosing, at every moment, which, if any, keys to hit.  Why are you doing this?  If you are taking notes for your own benefit, as an aid to memory, then, presumably, you believe that, at some later date, you will be able to see or hear or feel (depending on the device used for readout) something that will be informative about your prior choices of key-strokes.  If you are taking notes for someone else's benefit, then, presumably, you believe that, at a later date, that other agent will be able to see or hear or feel something that will be informative about your earlier choices of key-strokes.\footnote{One might also consider the following situation.  You are reading a text of which you are not the author.   Presumably, you take what you are reading as informative about what choices of keystrokes the author made while composing it.  If not, then why are you still reading?}

Why are you using a laptop, rather than, say,  a kumquat?   Presumably, you believe that there is something about the internal structure of the laptop that makes it suitable for the purpose. Suppose that you can use the readout to distinguish between two alternative choices of things to write today. For concreteness, consider a choice between writing
\begin{enumerate}[label=\Alph*]
\item `Twas brillig, and the slithy toves did gyre and gimble in the wabe.
\item Now is the winter of our discontent made glorious summer by this sun of York.
\end{enumerate}
If you believe that the readout tomorrow will allow you to distinguish between having made choice $A$ and choice $B$ today (in case you're forgetful)---that is, if you believe that the laptop can serve as an auxiliary memory---this is to attribute to the laptop a certain role in a causal process linking your choice today to your experience tomorrow.  Even if you have little or no idea of the internal workings of the laptop, attribution of this causal role to the laptop involves an assumption that your action will  have an effect on the internal state of the laptop, and the effect your action had on the internal state of the laptop will, in turn, have an effect on what it is that you experience tomorrow.

Now, an extreme version of operationalism might forbid  you to entertain conjectures about the internal state of the laptop, and speak only of the relations between your act today and the readout tomorrow, with no mention of intervening variables.  But the fact is that you are using a laptop rather than a kumquat, and, in fact, would not attempt to use a kumquat for this purpose, and that means that you believe that there is something about the physical constitution of the laptop that makes it suitable for the role you wish it to play in the causal chain between your act today and the readout tomorrow. Caution should  be exercised in theorizing about the internal workings of the laptop, but, as long as we are careful not to be committed to anything more than what we have good evidence for, there is no harm, and some potential benefit, in theorizing about the processes that mediate your actions today and experiences tomorrow.

Suppose, then, that you begin---tentatively, and with all due caution---to form a theory about the inner workings of the laptop. What constraints do your beliefs about the input-output relations of the laptop place on the sorts of theories you should entertain about its inner workings?

It seems that there are a few general things that can be said.   If you believe that the laptop can serve as a means for discriminating (with certainty) between act $A$ and act $B$, you should believe that there are distinct sets of internal states corresponding to these two  acts.  That is, the set of states that the laptop could end up in, as a result of your performing act $A$, has no overlap with the set of states that it could end up in as a result of your performing act $B$.  Moreover, these two sets of states are distinguishable upon readout---the readout mechanism produces outputs, corresponding to the two sets of internal states associated with acts $A$ and $B$, that are perceptually distinguishable by you. The only alternative to believing this is to believe in a direct causal effect of your choice today on your experience tomorrow, unmediated by any effect on the state of the world in the interim.

Of course, discrimination with absolute certainty is too much to ask.  You might not regard your laptop as a \emph{completely} reliable means for obtaining information about your past choices of key-strokes.  That's okay; it can still be informative, as long as if you take some perceived outputs to be more likely, given some choices of keystrokes, than others.  We can cash this out in terms of your conditional degrees of belief.

All of this will be formalized in the next section.

\section{Constructing a framework}\label{Fram}  Let us begin to formalize the informal considerations of the previous section.\footnote{The framework of this section is based on the ontological models framework of \citet{HarriganSpekkens}.}

To simplify matters, we will imagine a Bayesian agent, Alice, whose gradations of strength of belief in various propositions can be represented by a real-valued function satisfying the axioms of probability.  We will call this function the agent's \emph{credence function}, and the value that it takes on a given proposition $p$, the agent's \emph{credence} in $p$.  We will avoid talking of ``probability,'' in case that, for some, the word has connotations of something non-subjective.

We consider some physical system (you may think of the laptop as a running example), with which our agent will engage.  She has some set of operations $\{\phi_1, \ldots, \phi_n\}$ that can be performed on the system at time $t_0$ (think choices of key-strokes).  We will also consider situations in which an act is chosen without Alice's knowledge of which one; she has an $n$-sided die that she can toss (and when she does, her credence is equally divided among the possible outcomes), and a gadget that will choose among the operations $\{\phi_1, \ldots, \phi_n\}$, depending on the outcome of the die-throw.

 At some later time $t_1$, she performs an operation $R$ (a ``readout'' operation), which will result in one of a set of perceptually distinguishable results $\{r_1, \ldots r_m\}$.  (We can, of course, generalize, and give her a choice of operations to perform, but for our purposes, one will suffice.)

We assume that our agent has conditional credences, $cr^A(r_k | \phi_i)$, representing credence in result $r_k$ on the supposition of act $\phi_i$.  Upon learning the result of the operation $R$, she  uses these to update, via conditionalization, her credences about which act  was performed.

Alice does not believe that the acts $\{\phi_i\}$ have a direct influence on the outcomes $\{r_k\}$, unmediated by any effect of those acts on the system $S$.  We will permit her to entertain various theories about the workings of the system.  Any theory $T$ of that sort will involve a set $\Omega_T$ of possible physical states.  At the level of generality at which we are operating, we will assume nothing about the structure of these state-spaces  $\Omega_T$, and, in particular, do not assume that they are either classical or quantum in structure.  We do assume that, on the supposition of theory $T$, Alice can form credences in propositions about the state of the system $S$ to the effect that the system's state is in a subset $\Delta$ of  $\Omega_T$, and that, for each theory $T$, there is an algebra $\mc{A}_T$  of subsets of $\Omega_T$ deemed suitable for credences of that sort.

Suppose, now that  Alice regards the state of the system to be relevant to the outcome of the readout operation.  What she thinks about that relevance may, of course, depend on the theory $T$.  To represent these judgments, we suppose her to have conditional credences of the form  $cr^A(r_k | \phi_i, T, \omega)$.  This is her conditional credence, on the supposition of theory $T$ and the supposition that the state of the system is $\omega \in \Omega_T$,  that the result of the readout will be $r_k$, if act $\phi_i$ is performed.   It should be stressed that these are \emph{Alice's} conditional credences; there is no assumption that these conditional credences mirror any causal structure out there in the rest of the world.  Alice may hope that her conditional credences reflect something of the causal structure of the world (or rather, the causal structure, according to theory $T$), but  no assumption to the effect that they do  reflect the causal structure of the world will form any part of our argument, because our conclusions will only be about what Alice should believe.

We assume, also, that, on the supposition of $T$, Alice has credences about the states produced by the acts $\phi_i$. For any subset $\Delta$ of $\Omega_T$ that is in   $\mc{A}_T$,  let   $cr^A(\Delta | T, \phi_i)$ be Alice's conditional  credence, on the supposition of theory $T$, that performing act $\phi_i$ would put the system into a state in $\Delta$.  Her credences about the state should mesh with her conditional credences $cr^A(r_k | \phi_i, T, \omega)$ to yield her act-outcome conditional credences, on the supposition of theory $T$:
\begin{equation}\label{mesh}
cr^A(r_k | \phi_i, T) = \ex{cr^A(r_k | \phi_i, T, \omega)},
\end{equation}
where the angle-brackets $\ex{ \; \cdot \;}$ indicate expectation value taken with respect to her credences  $cr^A(\Delta | T, \phi_i)$ about the state, given act $\phi_i$.  If $\{ T_j \}$ is the set of theories that Alice takes seriously enough to endow with nonzero credence, then her act-outcome conditional credences $cr^A(r_k | \phi_i)$ should satisfy
\begin{equation}\label{expect}
cr^A(r_k | \phi_i) = \sum_j cr^A(r_k | \phi_i, T_j) \, cr^A(T_j).
\end{equation}

As mentioned, Alice does not believe that the acts $\{\phi_i \}$ have a direct influence on the outcomes $\{r_k \}$, unmediated by any effect of those acts on the system $S$.  This will be reflected in her conditional credences; her belief can be captured by the condition  that $cr^A(r_k | \phi_i, T, \omega )$ be independent of which act is performed.  That is, for all $i, j$,
\begin{equation}\label{screen}
cr^A(r_k | \phi_i, T,\omega )  = cr^A(r_k | \phi_j, T, \omega),
\end{equation}
for all states $\omega \in  \Omega_T$.  This is the condition that her choice of act has an influence on the later outcome \emph{only} via the  influence of this choice on the state of the system.

Now, part of the point of taking notes on a laptop is that what you see later will permit you to distinguish between choices of key-strokes you made earlier. There may be a limit to such discrimination; if you type something and then delete it, the later  readout might fail to distinguish between alternatives for the deleted text.  Some pairs of acts will be readout-distinguishable; others will not.   We will say that Alice takes two acts $\{\phi_i, \phi_j\}$ to be $R$-\emph{distinguishable} if, for every outcome $r_k$ of the operation $R$, either $cr^A(r_k | \phi_i)$ or  $cr^A(r_k | \phi_j)$ is zero.

If Alice takes two acts  to be $R$-distinguishable, then, provided she takes the influence of acts performed on the system on the later readout to be mediated by a change in the state of the system, then she should also take it that the sets of states that can result from the two acts are distinct.  We can formalize this; say that Alice takes the effects of acts $\{\phi_i, \phi_j\}$ to be \emph{ontically distinct}, on the supposition of theory $T$, if there exists  $\Delta \subseteq \mc{A}_T$ such that  $cr^A(\Delta | T, \phi_i) = 1$ and  $cr^A(\Delta | T, \phi_j) = 0$.    We have the following simple theorem.

\begin{theorem}
If an agent's  credences satisfy (\ref{mesh}),  (\ref{expect}) and (\ref{screen}), then, if she takes two acts to be $R$-distinguishable for some operation $R$, then she also takes the effects of those acts to be ontically distinct on any theory $T$ in which she places nonzero credence.
\end{theorem}
We emphasize: this is a theorem about Alice's credences, not about actual states of the system.  Alice could be wrong in her judgment that the effects on her laptop of different choices of keystrokes are ontically distinct.  It might be that her keystrokes have no effect whatsoever on the internal state of the laptop (in which case she would be misguided in taking those acts to be readout-distinguishable).

All of this is very general, and no assumptions have been made about the  sorts of theories that Alice might entertain. Now, suppose we apply it to quantum mechanics, with quantum probabilities construed subjectively.  Suppose that Alice's credences about the results of future readout operations performed on system $S$ can be represented by quantum states, and that which quantum state it is that represents her credences depends on her  choice of act.  If two acts are distinguishable by some experiment that can be done on the system, the corresponding quantum states are orthogonal.  If Alice places nonzero credences only in theories for which her conditional credences satisfy  (\ref{screen}), then she should also believe that the effects of acts with which she associates orthogonal quantum states are ontically distinct.

\section{Ontic distinctness of non-orthogonal states} We have come to the conclusion that Alice should take the effects of preparation-acts that she regards as $R$-distinguishable for some operation $R$ to be ontically distinct.  This includes acts with which she associates orthogonal quantum states. But what about non-orthogonal quantum states?

Let $\{\phi_1, \ldots, \phi_n \}$ be a set of preparation-acts that are such that Alice's credences about the results of future experiments, upon performance of these acts, can be represented by pure quantum states $\{ \ket{\phi_1}, \ldots, \ket{\phi_n} \}$.  If Alice takes the effects of these acts to be ontically distinct, this means that she believes that the physical state carries a trace of which preparation was performed and, that, if she knew the physical state, this would be enough to specify which of the quantum states $\{ \ket{\phi_1}, \ldots, \ket{\phi_n} \}$ she deems appropriate to use to set her credences about the results of future experiments.  That is, she believes that, for each physical state that could arise from one of these preparations, there is a unique quantum state that ``corresponds'' to it as the state to use in setting her credences.  The correspondence might not be one-one, as there might be a multiplicity of physical states that correspond to the same quantum state.

If Alice believes that the effects of any pair of preparation-acts with which she associates distinct pure quantum states are ontically distinct, then Alice believes that there is something in physical reality corresponding to these quantum states.  We will say, in such a case, that \emph{Alice is a realist about pure quantum states}.

The PBR theorem can be adapted to provide a set of conditions on Alice's credences sufficient for her to be a realist about pure quantum states.

Following the terminology of \citet{LeiferPsiOnt}, we will say that Alice takes a set $\{\phi_1, \ldots, \phi_n \}$ of preparation-acts to be \emph{antidistinguishable} if there is an operation $R$ such that, for each result $r_k$, $cr^A(r_k | \phi_i)$ is zero for at least one $\phi_i$.  Though Alice might not be able to uniquely decide which of the acts was performed, she will always be able to rule at least one out, whatever the result of operation $R$ is.

The setup of the PBR theorem is as follows.  Suppose that we have two systems, $S_1$ and $S_2$, of the same type, and suppose that each of them can be subjected to either  of two preparation-acts, $\{\phi, \psi \}$.  We assume that each choice of preparation act performed on  one system is compatible with both choices of act performed on the other. This gives us four possibilities of preparation-acts for the joint system, which we will designate by $\{{\phi_1 \otimes \phi_2},  {\phi_1 \otimes \psi_2}, {\psi_1 \otimes \phi_2}, {\psi_1 \otimes \psi_2} \}$.

Suppose, now, that Alice regards these four preparation-acts as antidistinguishable.  Under certain further conditions, which we will now specify, she should then take  the effects of the preparation-acts $\phi$ and  $\psi$  on the two systems to be ontically distinct.

A theory about the physical states of the two systems, presumably, would be able to treat $S_1$ and $S_2$ on their own, and also to regard the composite system that has $S_1$ and $S_2$ as parts as a system in its own right.  It will, therefore, include state-spaces $\Omega_1$ and $\Omega_2$ for the individual systems, and a state space $\Omega_{12}$ for the composite system.  In a classical theory, to specify the physical state of the composite system $S_1 + S_2$, it suffices to specify the states of the component parts, and so the state of the composite can be represented as  an ordered pair $\langle \omega_1, \omega_2 \rangle$ of states of the components, and $\Omega_{12}$ can be taken to be the set of all such ordered pairs. That is, $\Omega_{12}$ can be  taken to be the cartesian product of  $\Omega_1$ and $\Omega_2$.  For other theories, such as quantum mechanical theories, the cartesian product $\Omega_1 \times \Omega_2$ might be included in $\Omega_{12}$ without exhausting it.  In quantum mechanics, for any states $\psi_1, \psi_2$ of two disjoint systems, there is a corresponding product state $\psi_1 \otimes \psi_2$ of the composite,  but not all pure states of the composite system are of this form, as there also pure entangled states.

We will say that a theory $T$ satisfies the \emph{Cartesian Product Assumption} (CPA) if  $\Omega_1 \times \Omega_2$ is a subset of $\Omega_{12}$.  Given a theory satisfying the CPA, Alice's credences regarding a pair of preparations $\phi_1$, $\phi_2$ that can be performed on the two subsystems will be said to satisfy the \emph{No Correlations Assumption} (NCA) if her credences about the joint state of the two systems, on the supposition of $T$ and the preparations $\phi_1$, $\phi_2$, are concentrated on the cartesian product $\Omega_1 \times \Omega_2$ and if her credences are such that information about the state of one system would be completely uninformative about the state of the other.  The conjunction of the Cartesian Product Assumption and the No Correlations Assumption is called the \emph{Preparation Independence Postulate} (PIP).  Note that this is a combination of an assumption about the structure of the state space of a theory $T$ and an  assumption about Alice's credences, conditional on the supposition of $T$ and on the preparations $\phi_1$, $\phi_2$.

Now suppose that we have a pair of systems $S_1$, $S_2$, a theory $T$   about them that satisfies  the CPA, and, for each system, a pair of preparation-acts $\phi_i$, $\psi_i$, which are such that:
\begin{enumerate}
\item There is an operation $R$ that can be performed on the pair of systems, such that, for each outcome $r_k$ of the operation, at least one of\
\[
\{ cr^A(r_k | \phi_1 \otimes \phi_2), cr^A(r_k | \phi_1 \otimes \psi_2), cr^A(r_k | \psi_1 \otimes \phi_2), cr^A(r_k | \psi_1 \otimes \psi_2) \}
\]
is equal to zero.  That is, Alice regards these preparation-acts as antidistinguishable.
\item Alice's credences about the states of the systems $S_1$ and $S_2$, conditional on the supposition of theory $T$ and each of the preparations mentioned, satisfy the NCA.
\end{enumerate}
It then follows, by the argument of \citet{PBR}, that Alice should regard the effects of these preparations to be ontically distinct, conditional on the supposition of theory $T$.

Now, of course, Alice might entertain theories for which the CPA does not hold, and, among those for which it does, her conditional credences might satisfy the NCA for some but not for others.  In this case, she should not be certain that the effects of these preparations are ontically distinct.  However, her credence that they are must be at least as high as her total credence in the class of theories for which the above-listed conditions are true.  Another way to put this is: in order to be \emph{certain} that the effects of these preparations are not ontically distinct---that is, in order to categorically deny onticity of quantum states---Alice's credences in all such theories must be strictly zero.

Though it seems natural from a classical standpoint, quantum mechanics gives us incentive to question the Cartesian Product Assumption.  In the set of pure states of a bipartite system, product states are rather special, and every neighbourhood of any product state contains infinitely many entangled states.  If both systems have Hilbert spaces of infinite dimension, then the entangled states are dense in the set of all states, pure or mixed (see \citealt{CliftonHalvorson2000,EntangOpenSystems}).  We should, therefore, take seriously the cases of theories whose state spaces either lack a Cartesian product component, or are such that product states cannot reliably be prepared exactly, although  we may be able to approximate them arbitrarily closely.  A modification of the PBR argument that does not employ the Cartesian Product Assumption is therefore desirable.

In \citet{MyrvoldPsiOnt,MyrvoldQR}, a substitute for the PIP is proposed, which is strictly weaker than it and dispenses with the Cartesian Product Assumption.  This substitute is called the Preparation Uninformativeness Condition (PUC).

 To state the assumption, we consider the following set-up.  Suppose that, for systems $S_1$, $S_2$, we have some set of possible preparation-acts that can be performed on the individual systems.  Suppose that the choice of preparation for each of the subsystems is made independently, say, by rolling two separate dice.  Following the preparation of the joint system, which consists of individual preparations on the subsystems, you are not told which preparations  have been  performed, but you are  given a specification of the ontic state of the joint system.  On the basis of this information, you form credences about which preparations were performed. In the case of preparations whose effects you take to be  ontically distinct, you will be certain about what preparation has been performed; otherwise, you may have less than total information about which preparations were performed.

We ask: under these conditions, if you are now given information about which preparation was performed on one system, is this  informative about which preparation was performed on the other? The Preparation Uninformativeness Condition is the condition that it is not.  This condition is satisfied whenever the PIP is.  It is also satisfied if you regard the effects of the preparations to be ontically distinct: in such a case, given the ontic state of the joint system, you are already certain about which preparations have been performed, and being told about the preparation on one system will not shift your credences.

The PUC is implied by the PIP, but it is strictly weaker.  Even if the CPA is assumed, it is possible to construct models for the PBR setup, outlined in the previous section, in which the PUC is satisfied but the PIP is not. See \citet{MyrvoldPsiOnt} for one such construction.

On the assumption of the Preparation Uninformativeness Condition---which, in the current context, is a condition on the agent's credences, independent of the structure of the state space
of the theory considered---one can get a $\psi$-ontology proof for a class of nonorthogonal states. Let $\{\phi_i, \psi_i\}$ ($i = 1, 2$) be preparation-acts such that Alice's credences
about the results of future experiments can be represented by quantum states $\{ \ket{\phi_i}, \ket{\psi_i}\}$.   If ${|\bkt{\phi_i}{\psi_i}| \leq 1/\sqrt{2}}$, then $\{\ket{\phi_1}\ket{\phi_2},  \ket{\phi_1}\ket{\psi_2}, \ket{\psi_1}\ket{\phi_2}, \ket{\psi_1}\ket{\psi_2}\}$ is an antidistinguishable set.
From these assumptions, together with a mild side-assumption called the \emph{principle of extendibility}\footnote{This says that any system
composed of $N$ subsystems of the same type can be regarded
as a part of a larger system consisting of a greater number of
systems of the same type.}, it  follows  that, for any theory $T$ such that Alice's credences satisfy the PUC  for each of the preparation-acts
 $\{ \phi_1 \otimes \phi_2, \phi_1 \otimes \psi_2, \psi_1 \otimes \phi_2, \psi_1 \otimes \psi_2 \}$, Alice  takes the effects of the preparation acts $\{\phi_i, \psi_i\}$ to be ontically distinct.

A key feature of quantum mechanics used in these proofs is the fact that, if $\{\ket{\phi_i}, \ket{\psi_i} \}$ are any state vectors with $|\bkt{\phi_i}{\psi_i}| \leq 1/\sqrt{2}$, then
\[\{{\ket{\phi_1}\ket{\phi_2},}  {\ket{\phi_1}\ket{\psi_2},} {\ket{\psi_1}\ket{\phi_2},} {\ket{\psi_1}\ket{\psi_2}}\}\] is an antidistinguishable set.  What is needed for the argument is that there exist some preparations and an operation such that Alice's preparation-response conditional credences match the Born-rule probabilities yielded by these states for the outcomes of the experiment envisaged.  This is not a matter of either logical necessity or probabilistic coherence.  What we will assume is that Alice accepts quantum mechanics, in the following sense: for any quantum state $\rho$ of a system, and any complete set of mutually orthogonal projections $\{P_1, \ldots, P_n\}$ on the Hilbert space of the system, there is \emph{some} combination of preparation-act and readout-operation such that Alice's credences in the possible results $\{ r_1,\ldots, r_n\}$ match (or at least closely approximate) the Born-rule probabilities $\mbox{Tr}(\rho \, P_k)$.\footnote{Of course, we could reasonably expect something stronger, that this holds, not just for projections, but also for any positive-operator valued measure (POVM).  But the weaker condition is all that we need.}

The upshot of these theorems is that, for any theory $T$ for which the assumptions (which may include both assumptions about the theory and about Alice's credences conditional on the supposition of the theory), Alice's credences, conditional on $T$, should take the effects of the preparation-acts considered to be ontically distinct.  Thus, her credence in the ontic view of quantum states should be at least as high as her total credence in the class of all such theories.  The only way that she can categorically deny that preparation-acts corresponding to distinct quantum states yield  ontically distinct states of reality is to attach credence zero to the class of all such theories.

\section{The QBist response}  Advocates of QBism will not accept the conclusion that a subjectivist about quantum probabilities should be a realist about  quantum states. Yet the argument so far has not violated any of the core tenets of QBism.  To avoid the conclusion,  QBists must add to its main tenets a further prohibition, logically independent of the explicitly stated core tenets.

In section 2, we first  argued for what, in my opinion, ought to be an uncontentious claim: that an agent who regards a pair of preparation-acts to be distinguishable ought to regard the  effects of these acts to be ontically distinct. This has the consequence that, if an agent associates orthogonal quantum states with a pair of preparation-acts, she ought to take their effects to be ontically distinct.  The extension of this conclusion to nonorthogonal quantum states requires additional assumptions, which might be questioned.  But the QBist will want to resist the very first step, having to do with orthogonal quantum states.

Nonetheless, the reasoning involved does not violate what proponents of QBism have identified as its fundamental tenets.  These are given in different versions in different venues, but  this is true for any of these versions.

Fuchs, Schack, and Mermin (\citeyear{FMSQB}) present the following as the ``three fundamental precepts'' of QBism.
\begin{enumerate}
\item A measurement outcome does not preexist the measurement. An outcome is created for the agent who takes the measurement action only when it enters the experience of that agent. The outcome of the measurement is that experience. Experiences do not exist prior to being experienced.
\item	An agent's assignment of probability 1 to an event expresses that agent's personal belief that the event is certain to happen. It does not imply the existence of an objective mechanism that brings about the event. Even probability-1 judgments are judgments. They are judgments in which the judging agent is supremely confident.
\item Parameters that do not appear in the quantum theory and correspond to nothing in the experience of any potential agent can play no role in the interpretation of quantum mechanics.
\end{enumerate}

Regarding Precept 1: We have not assumed that the result of a readout-operation (we have refrained from using the misleading term ``measurement'') preexists the operation, or that experiences exist prior to being experienced.  Nor have we presupposed that the result of such an operation merely indicates a preexisting element of reality.

Regarding Precept 2: What we have concluded is that, if an agent takes two preparation-acts to be distinguishable with certainty, the agent should take the effects of those acts to be ontically distinct.  The agent may be mistaken about this, of course.  For example: I may be convinced that I will be able to learn  tomorrow about  my choices of key-strokes today, because I regard different choices of what to write today as readout-distinguishable tomorrow.  Accompanying this belief is  a belief that distinct choices of keystrokes today, if I save the file and don't erase it, have the effect of putting my laptop's internal mechanism into physically distinct states, which will have observationally distinguishable effects tomorrow.  I could be wrong about this, of course.  It might be that my laptop is malfunctioning and that there are, in fact, no lasting effects of my choices of keystrokes today. But---and this is the crucial point---if I were to revise my judgment about the ontic effects of my keystrokes, I would also revise my assessment of the readout-distinguishability of my choices tomorrow.  What I would \emph{not} do is simultaneously maintain that my  choices today have no effect on the internal state of the laptop and that the result of the readout operation tomorrow will be informative about those choices.

Regarding Precept 3: The conclusion that an agent who takes  two preparation acts to be distinguishable  should regard the effects of these acts to be ontically distinct is quite general, and is not specific to quantum mechanics. It invokes the sort of considerations that can be brought to bear on any sort of theorizing about the world.  It should also be emphasized that the mere introduction of some sort of physical state-space $\Omega$ does not automatically consign us to the realm of hidden-variables interpretations of quantum mechanics.  The structure of the physical state-space is left completely open, and our conclusion that one should take there to be distinctions in physical reality corresponding to distinctions between pure quantum state-preparations does not, by itself, commit us to anything else.

\citet{FuchsNotwithstanding} presents the ``three tenets'' of QBism as
\begin{enumerate}
\item The Born rule is a normative statement.
\item All probabilities, including all quantum probabilities, are so subjective they never tell nature what to do. This includes probability-1 assignments. Quantum states thus have no ``ontic hold'' on the world.
\item Quantum measurement outcomes just are personal experiences for the agent gambling on them.
\end{enumerate}
Regarding Tenet 1: Neither the Born rule, nor anything at all quantum-mechanical, was invoked in the argument for the conclusion that an  agent should take the effects of  preparation-acts that she regards as readout distinguishable to be ontically distinct. A special case of this is a pair of preparations that are such that her
credences can be represented by a orthogonal quantum states.  In the extension to non-orthogonal quantum states, Born-rule probabilities were invoked only in consideration of what sorts of credences an agent who accepts quantum mechanics should have.

Regarding Tenet 2: Again, we have not assumed anything at all about what nature does.  Our conclusion has only to do with the meshing of an agent's act-outcome conditional credences with her credences about the states of the physical system that mediates between the acts and the outcomes.  As we have already mentioned in connection with the malfunctioning laptop, that system is under no obligation to conform to her credences, and, she may, indeed, find her expectations disappointed. If the last sentence in the statement  of Precept 2  is supposed to indicate a denial of an ontic construal of quantum states, then, despite the ``thus,'' it
does not follow from the first. As we have seen, one can begin with the assumption that quantum probabilities are subjective and end with a conclusion that an agent whose credences conform to quantum mechanics ought to take quantum states to be ontic.

Regarding Tenet 3: Nothing in the argument depends on whether the results $\{ r_k \}$ of the operation are taken to be things like states of the laptop screen, or states of the agent's nervous system, or her conscious experience upon looking at the screen.

Given that we have not, in our arguments, violated the fundamental precepts or tenets of QBism, why would, and how could, a QBist resist the conclusion?  The only option open is  to reject the claim that there is a  causal relation between my choices of preparation-acts today and my experiences of the results of readout operations tomorrow that  is mediated by the effects of my actions on the state of the physical system on which I act. There are two ways to do this. One is to maintain that there is a direct causal link between my present actions and future experiences, unmediated by any effect of my actions on the physical system on which I act.  The other is to reject all talk of causal links between my present actions and future experiences.

There are passages in some writings by QBists that suggest the latter.
\begin{quote}
There is a sense in which this unhinging of the Born Rule from being
a ``law of nature'' in the usual conception \ldots i.e., treating it as a normative statement, rather than a descriptive one --- makes the QBist notion of quantum \emph{indeterminism} a far more radical
variety than anything proposed in the quantum debate before. It says in
the end that nature does what it wants, without a mechanism underneath,
and without any ``hidden hand'' (Fine 1989) of the likes of Richard von
Mises's \emph{Kollektiv} or Karl Popper's \emph{propensities} or David Lewis's objective
chances, or indeed any conception that would diminish the autonomy of
nature's events. Nature and its parts do what they want, and we as
free-willed agents do what we can get away with. Quantum theory, on
this account, is our best means yet for hitching a ride with the universe's
pervasive creativity and doing what we can to contribute to it \citep[272]{FuchsNotwithstanding}.
\end{quote}
The suggestion is that the physical world obeys neither deterministic nor stochastic laws.

What sort of physical laws, if any, nature conforms to is not something that can be known \emph{a priori}.  If someone proposes a physical theory according to which one should expect to observe certain statistical regularities, that theory can be subjected to empirical test according to the usual Bayesian methods. This will never produce certainty that the theory is correct, but can result in high credence that the theory, or something like it, is at least approximately correct within  a certain domain of application.  In particular, if someone were to propose a theory according to which a given sort of preparation-act (say, passing a particle of a certain type through a Stern-Gerlach apparatus oriented vertically, and selecting the $+$ beam) bestowed determinate chances on certain sorts of subsequent experiments, we could test that assertion by repeating the preparation-experiment sequence, and looking to see whether the relative frequencies converged to a stable value, in accordance with the weak law of large numbers.  If we routinely find  the theoretical expectation fulfilled, this should boost credence in the theory; if the outcomes of the experiment are not at all like what the theory leads us to expect, this counts as evidence against the theory.   One could imagine that persistent failure to find \emph{any} sort of regularity at all might boost credence in the proposition that there are no regularities to be found---if the agent the agent could survive that long!  (It may be a requirement for the existence of agents that could have credences that there be some modicum of predictability in the world.)

Despite passages like the one quoted above, I am skeptical that any QBist actually rejects the claim that there is a causal link between their present actions and their future experiences, mediated by an effect of their actions on the physical systems with which they interact.  I find it difficult to imagine how someone whose beliefs were like that would act. I am certain, however, that someone with beliefs like that  would not judge  a laptop that was advertised as having more memory---that is, more readout-distinguishable internal states that the user could choose between via choice of actions performed on the machine---to be worth more money than one with less memory.  Indeed, it is hard to see why an agent like that could prefer any laptop at all to a kumquat as an implement for taking notes.\footnote{It is worth re-emphasizing: I am \emph{not} claiming that anything about the laptop or the kumquat follows from assertions about the agent's credences about these systems.  The agent may prefer the laptop to the kumquat as a note-taking device because of a \emph{false} belief that the laptop is more suitable to play that role in a causal chain between the agent's actions and future experiences. The point is merely that the agent's belief, true or false, is a belief about the physical workings of the laptop.}

Fuchs does not explain what, if anything, he takes to be evidence for the radical indeterminism of which he speaks.  Such evidence \emph{cannot} have come from the results of quantum experiments, which, on the contrary, seem to indicate law-like connections, best expressed in terms of probabilistic laws, between preparation-acts and experimental outcomes.    At any rate, the evidence \emph{against} the claim that the world is so completely devoid of law-like regularities that theorizing about it is pointless seems to be overwhelming.

There is, it seems, one possible response for the QBist. This  would be to say that  he accepts the thesis of radical indeterminism, of a world subject neither to deterministic nor stochastic laws, not on the basis of evidence, but as a leap of faith, or an expression of temperament (see \citealt[251-253]{FuchsNotwithstanding}).  Though he believes in no causal connection between his actions on the world and future experiences, as a psychological weakness he is unable to set his credences about the future at will, and can only affect his credences about the future by undergoing a ritual manipulation of the physical objects around him.  He finds himself in the position of someone who does not believe that a horseshoe hung on the door will bring good luck, but hangs it anyway because he finds that doing so makes him more optimistic about the future.  If that is the view---namely, that a QBist can offer no reason whatsoever to accept  the radical renunciation of physical theorizing about the world that acceptance of the view entails---then we are in full agreement on that point.

\section{An argument from locality?} In addition to the tenets and precepts listed above, QBism also involves a severe restriction on the application of quantum mechanics.  An agent is required to only apply quantum probabilities to her own future experiences, and not to events (including the experiences of other agents) that she herself will not experience.

The severity of this restriction of scope is not always sufficiently emphasized.  Though some of the ideas adopted by QBists are inspired by quantum information theory, the restriction of credences to one's own experiences eliminates vast  swaths of quantum information theory, including everything to do with communication and cryptography.   The point of a communication is to influence the credences of another agent, and, in designing a communication protocol, one must  consider probabilities of changes to the recipient's belief-state.\footnote{It might be claimed, in response, that Alice, in sending a signal to Bob, is concerned about the impact on Bob only insofar as it will later impinge on her.  This, as everyone knows, is false.  Some of the intended results of a communication are ones that the sender will never be aware of.  As a particularly vivid example, consider a spy whose cover has been blown, who sends out one last encrypted message before taking a cyanide pill to avoid capture. A QBist must regard this as a misuse of quantum mechanics if the encryption scheme is a quantum scheme.  It is not!  Nor is it a misuse of quantum mechanics for an engineer designing a nuclear waste storage facility to use quantum mechanics to compute half-lives with the aim of constructing a facility that will be safe for generations to come.}

There may seem to be an  advantage to this restriction: one thereby avoids quantum nonlocality.  If an agent uses quantum mechanics to assign probabilities to events along her own worldline, and never to events at a distance from herself, all the events to which she assigns probabilities are timelike-related.  There can be no question of action-at-a-distance because there are no distances between the events she assigns credences to.

This advantage obtains only if one refrains from doing any theorizing, quantum or otherwise, about events at a distance.  Any theory, quantum or otherwise, that attempts to account for outcomes of tests of Bell inequalities---if we mean by `outcomes' what is usually meant, namely,  detector-registrations that occur at spacelike separation from each other---will have to violate at least one of any set of conditions sufficient to yield Bell inequalities.

The seeming advantage is illusory.  If one refrains from giving an account of goings-on in the world that occur at a distance from each other \emph{in order to avoid giving an account of nonlocal goings-on}, and if it is held that such a renunciation is \emph{necessary} to avoid some sort of  nonlocality that is thought to be objectionable, this is tantamount to saying that \emph{if one were to give an account of goings-on in the world that occur at a distance from each other, such an account would involve the objectionable sort of nonlocality}.   If this differs from acknowledging that the supposedly objectionable nonlocality is a feature of the world, I don't know how. It is as if one declined a request to evaluate  a co-worker by saying that one is declining the request in order to avoid saying that the co-worker  is incompetent.  In giving such a response, one has not refrained from conveying one's judgment of  the co-worker's competence.

\section{Conclusion}  Subjectivism about probabilities is not, by itself, an escape from $\psi$-ontology theorems.  The theorems go through with all probabilities interpreted subjectively, with a conclusion that an agent whose credences satisfy the conditions of the proof should believe that preparation procedures that she associates with distinct pure quantum states leave the systems on which those procedures are performed in distinct ontic states.  If this conclusion is to be avoided, some other stricture must be placed on the agent's credences.

The chief condition that underlies the proof is the assumption of the possibility of performing a preparation procedure that effectively screens off correlations between the system on which it is performed and the rest of the world, rendering a further specification of its physical state uninformative about the states of other systems.  This sort of assumption lies  deep at the heart of virtually all experimental science.   None of the empirical evidence that motivates a shift from classical to quantum theory provides grounds for rejecting assumptions of this sort. If we ever come to a point at which we have reasons to doubt this sort of assumption, this will not come about as a result of experiments that presuppose it.  And if we were presented with a reason to doubt this sort of assumption, it is hard to see how this doubt could be sufficiently contained so as not to undermine all of experimental science.  Fortunately, we are not in such a position.

\newpage

\bibliographystyle{chicago}

\end{document}